\documentclass[aps,pra,superscriptaddress,twocolumn,showpacs]{revtex4}
\usepackage{graphicx}
\usepackage{amsmath}
\usepackage{amsfonts}
\usepackage{epstopdf}
\usepackage{euscript}
\usepackage{subfigure}
\usepackage[usenames]{color}

\begin{document}
\title{Dissociative electron attachment to carbon dioxide via the $^2\Pi_u$ shape resonance}
\author{A.~Moradmand}
\affiliation{Department of Physics, Auburn University, Auburn, Alabama 36849, USA}
\author{D.~S.~Slaughter}
\affiliation{Lawrence Berkeley National Laboratory, Chemical Sciences, Berkeley, California 94720, USA}
\author{D.~J.~Haxton}
\affiliation{Lawrence Berkeley National Laboratory, Chemical Sciences, Berkeley, California 94720, USA}
\author{T.~N.~Rescigno}
\affiliation{Lawrence Berkeley National Laboratory, Chemical Sciences, Berkeley, California 94720, USA}
\author{C.~W.~McCurdy}
\affiliation{Lawrence Berkeley National Laboratory, Chemical Sciences, Berkeley, California 94720, USA}
\affiliation{Departments of  Chemistry and Applied Science, University of California, Davis,  CA 95616}
\author{Th.~Weber}
\affiliation{Lawrence Berkeley National Laboratory, Chemical Sciences, Berkeley, California 94720, USA}
\author{S.~Matsika}
\affiliation{Department of Chemistry, Temple University, Philadelphia, Pennsylvania 19122, USA}
\author{A.~L.~Landers}
\affiliation{Department of Physics, Auburn University, Auburn, Alabama 36849, USA}
\author{A.~Belkacem}
\affiliation{Lawrence Berkeley National Laboratory, Chemical Sciences, Berkeley, California 94720, USA}
\author{M.~Fogle}
\affiliation{Department of Physics, Auburn University, Auburn, Alabama 36849, USA}

\date{\today}

\begin{abstract}
Momentum imaging measurements from two experiments are presented and interpreted with the aid of new {\em ab initio} theoretical calculations to describe the dissociative electron attachment (DEA) dynamics of CO$_2$. The dynamics of the transient negative ions of CO$_2^-$ involve several conical intersections taking part in mechanisms that have only recently been understood. We address the problem of how the 4~eV $^2\Pi_u$ shape resonance in CO$_2$ proceeds to dissociate to CO($^1\Sigma^+$) + O$^-$($^2$P) by DEA. 
\end{abstract}

\maketitle
\section{Introduction}
Low-energy electron scattering from CO$_2$ is dominated by the well-known $^2\Pi_u$ shape resonance\cite{Bone1969,Morrison1977,allan_selectivity_2001,resc2002,mccurdy_resonant_2003} at 4~eV collision energy and a dramatic rise in the total cross section below 1 eV which has been attributed to a $^2\Sigma^+$ virtual state\cite{vanr2004,allan_vibrational_2002,somm2004}.  A temporary negative ion (TNI) CO$_2^-$ is also formed at 8.2~eV due to a $^2\Pi_g$ electronic Feshbach resonance that can decay either by electron autodetatchment or dissociation into O$^-$ and vibrationally hot CO in its electronic ground state, $^1\Sigma^+$~\cite{chant1972,sizun1979,dressler_energy_1985,slaughter_dissociative_2011,wu_renner-teller_2012}.  The question we address in paper is how the  $^2\Pi_u$ shape resonance state at 4 eV dissociates to O$^-$ + CO in their ground electronic states, since it is the higher  $^2\Pi_g$  Feshbach resonance that correlates directly to those products.

 The absolute cross section for DEA measured at the peaks of the 4~eV and 8.2~eV resonances reported in the literature~\cite{stamatovic1973,Orie1983} are $1.5\times10^{-19}$ and $4.5\times10^{-19}$ cm$^2$, respectively. Dissociation through the 8.2 eV resonance results in a bimodal translational kinetic energy distribution of the two products, reflecting two distinct CO vibrational distributions. Similarly, the 4~eV $^2\Pi_u$ shape resonance decays by autodetachment, with significant accompanying vibrational excitation, or dissociation into O$^-$($^2$P) and CO($^1\Sigma^+$). By symmetry, there are only three electronic states that  can correlate to the CO($^1\Sigma^+$) + O$^-$($^2$P) asymptote, corresponding to the three possible projections of the oxygen P state. In linear geometry, these states consist of a doubly degenerate $^2\Pi$ state and a $^2\Sigma$ state that, as O$^-$ approaches CO, becomes a virtual state.
The correlation between the final states and the 4- and 8.2~eV resonance states was recently found~\cite{slaughter_dissociative_2011} in the linear geometry, where the doubly excited 8.2 eV $^2\Pi_g$ state feeds the O$^-$($^2$P) and CO($^1\Sigma^+$) dissociation limit, while the 4 eV $^2\Pi_u$ state dissociates to O($^1D$) and a short lived CO$^-$($^2\Pi$) anion, via an avoided crossing with the $^2\Pi_g$ state.  There are conical intersections among the $^2\Pi$ states close to the point where they avoid in linear geometry.

Conical intersections are known to play an important role in the dynamics of dissociative electron attachment (DEA), with one recently being identified between the $^2B_2$ and $^2A_1$ states of dissociating H$_2$O$^-$\cite{haxton_observation_2011}. In an effort to understand the dynamics that enable the CO$_{2}^{-}$($^2\Pi_u$) state to dissociate to O$^-$($^2$P) and CO($^1\Sigma^+$), we present results from two independent experiments that elucidate the kinetic energy and angular distributions of the final state products. In addition, we have performed {\em ab initio} theoretical calculations that enable a comparison of the laboratory-frame measurements with entrance amplitudes calculated in the molecular frame~\cite{hax2006} and give additional insight into the dynamics of dissociation in this system.

\section{Theoretical Considerations}
On the theoretical side, DEA to CO$_2$ via the 4 eV $^2\Pi_u$ shape resonance poses a number of challenges. 4eV is close to the thermodynamic limit for producing O$^-$ + CO. But as we recently mentioned~\cite{slaughter_dissociative_2011}, the CO$_{2}^{-}$($^2\Pi_u$) state correlates with O*($^1$D) + CO$^-$, which means DEA must necessarily involve a conical intersection with another anion state. We had previously identified a conical intersection between the $^2\Pi_u$ state and a doubly excited (Feshbach) $^2\Pi_g$ resonance which, in linear geometry, lies energetically above 4 eV and was therefore assumed not to play a role in the DEA process through the lower resonance~\cite{slaughter_dissociative_2011}. 

The observed angular distributions also pose a set of puzzles. The fact that the observed distributions are markedly different from what is expected for a $^2\Pi_u$ state implies that the transient negative ion undergoes bending as it fragments, signaling a strong breakdown of the axial recoil approximation. But we also know from earlier theoretical work~\cite{resc2002, mccurdy_resonant_2003} that in the vicinity of the neutral CO$_2$ geometry, the width of the lower component of the $^2\Pi_u$ state steeply increases upon bending, leading to rapid electron detachment. The implications of these facts point to a complex dissociation mechanism, involving a combination of stretching and subsequent bending along with non-adiabatic transitions. This leads us here to consider a conical intersection between the shape- and Feshbach states at a different geometry and also a conical intersection between the $^2\Pi_u$ state and the $^2\Sigma_g$ virtual state~\cite{somm2004}, since the latter also correlates with O$^-$($^2$P) + CO($^1\Sigma^+$).

\section{Experimental}
The Auburn University (AU) experiment consists of a negative-ion momentum spectrometer with a pulsed electron beam and supersonic gas jet in a crossed-beam geometry \cite{moradmand_momentum-imaging_2013}. The gas jet produces target molecules of approximately 15~K by adiabatically expanding room temperature CO$_2$ through a 10~$\mu$m aperture. The central portion of the resulting gas expansion is then selected by a 0.3~mm diameter skimmer to produce a molecular beam which is spatially confined in the interaction region. The electron energy resolution is estimated to be $\pm$0.5~eV  by comparing results from the 6.5~eV O$^-$ resonance in O$_2$ to well-established thresholds from literature as well as the appearance of the O$^-$ resonance at 4~eV in CO$_2$. A 4$\pi$ detection solid angle of the anions is achieved by pulsing a uniform electric field to extract any anion products. This extraction field is delayed to allow the electron pulse to clear the interaction region. The pulsing scheme is operated at a 40~kHz repetition rate. the extracted anions are detected by an 80~mm microchannel plate detector with a delay-line anode for time and position information. By using the time-of-flight and position information, a 3-dimensional momentum image can be determined for each anion product and can be compared to the initial electron momentum vector of angular correlation and kinetic energy determination.

The experimental technique for the Lawrence Berkeley National Laboratory (LBNL) experiments has been described in detail previously\cite{adaniya_momentum_2012}, so we will provide only a brief overview here. The momentum of the final state anionic fragment following DEA to single CO$_2$ molecules were measured using a momentum imaging negative ion spectrometer. A gaseous target beam was effused from a narrow stainless steel capillary to intersect with a magnetically-collimated electron beam that was pulsed with a 50~kHz repetition rate. The electron energy spread was typically 0.8~eV full width at half maximum (FWHM), measured using the slope of the DEA thresholds for O$^-$ production from CO$_2$ at 4.0~eV and H$^-$ production from H$_2$O at 6.0~eV, both of which were also used as a reference for the electron energy scale. Ion focusing optics within the ion spectrometer allowed the effective interaction volume to be reduced by a factor of 3. Ion momentum images were recorded from a position- and time-sensitive detector to an event list, so that 3-dimensional momentum distributions could be determined for a full $4\pi$ solid angle of detection.

\section{Results}
\begin{figure}
\begin{center}
\includegraphics[width=0.6\columnwidth,clip=true]{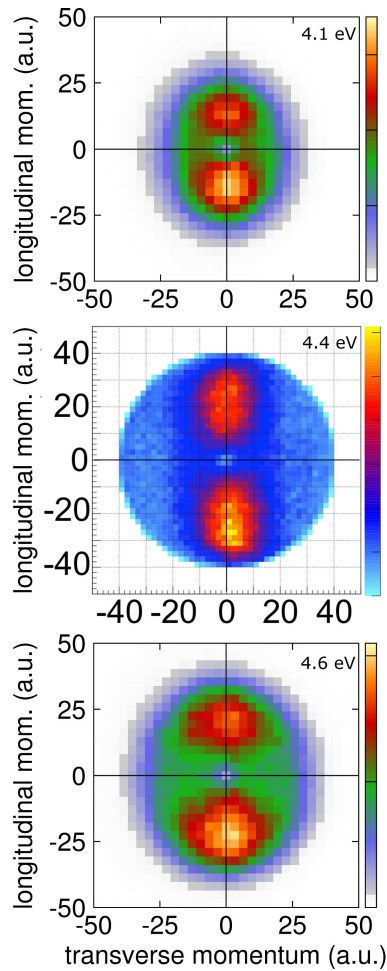}
\caption{(color online). Laboratory-frame O$^-$ momentum distribution following DEA to CO$_2$ at 4.1~eV nominal electron beam energy (a), 4.6~eV (c), measured using the LBNL apparatus. The 4.4~eV data (b) was measured using the Auburn apparatus on a colder CO$_2$ target that is better spatially confined, resulting in a higher momentum resolution. In each plot the incident electron direction is longitudinal, pointing upwards. The color-scales are linear and represent ion yield in arbitrary units.}\label{O-mom}
\end{center}
\end{figure}

The measured O$^-$ momentum distributions for three nominal electron beam energies are shown for the two experiments in Fig.\ref{O-mom}. 
In displaying the 3-dimensional momentum distribution on two axes, we have sliced the experimental data in momentum space so that the volume of momentum space contained in each bin is equivalent. Specifically, we have applied a gate on the data that is symmetric about a plane including the electron beam axis and increases in volume linearly with absolute momentum. The maximum of the ion momentum distribution near the 4.0~eV thermodynamic threshold for dissociation (Fig.~\ref{O-mom}a) occurs at 15~atomic units (a.u.) in the longitudinal direction and increases to 25~a.u. as the electron energy is increased above threshold (Figs~\ref{O-mom}b and c), while the ion momentum in the transverse direction remains significantly smaller in the LBNL data (Figs~\ref{O-mom}a and c) and negligible in the AU data (Fig.~\ref{O-mom}b). Both experiments found the O$^-$ momentum to be preferentially directed along the electron beam direction and the AU experimental data, measured with higher momentum resolution due to smaller spatial volume and lower temperature of the target, determined the breakup to be even more confined to the incident electron beam axis.

\begin{figure}
\begin{center}
\includegraphics[width=1.0\columnwidth,clip=true]{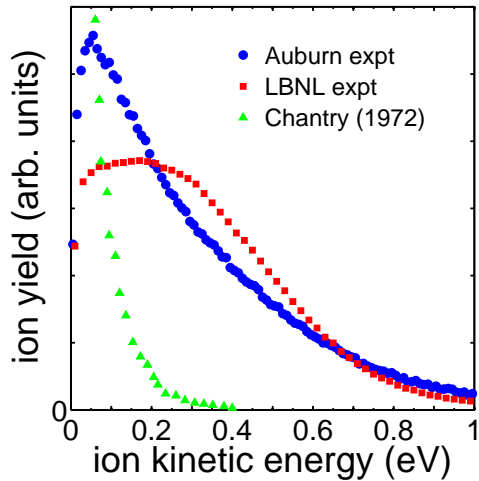}
\caption{(color online). Normalized O$^-$ kinetic energy distribution following DEA to CO$_2$, measured at LBNL and AU, at 4.4~eV nominal electron beam energy, respectively. The statistical uncertainties (one standard deviation) are smaller than the size of the plot markers. The data of \textcite{chant1972} (4.3~eV) are displayed for comparison.}\label{O-KE}
\end{center}
\end{figure}

The kinetic energy spectra of O$^-$ for the 4.4~eV AU and 4.6~eV LBNL experiments are displayed in Fig.~\ref{O-KE}. 
The differing widths of the curves between the two present experiments may be due to the different target temperatures and interaction volumes in each setup. In the LBNL experiment the CO$_2$ target is prepared at room temperature and is internally cooled to about 40~K\cite{ramsey_molecular_1985}, in the plane perpendicular to the beam propagation, by collimation through a capillary; however the broadening of the measured kinetic energy due to this finite target temperature is well-known and was modeled by \textcite{chan1967}. In the present LBNL data we estimate the broadening due to temperature to be typically less than 0.1~eV FWHM and less than 0.05~eV FWHM in the AU experiment, where the target jet was cooled to $\sim$15~K by supersonic expansion. The electron beam energy distribution, which is 0.8~eV FWHM and 1.0~eV FWHM in the LBNL and AU experiments, respectively, also contributes to the width of the measured ion kinetic energy spectrum. The data of \textcite{chant1972} are much more strongly peaked at low-kinetic energies. This could be due to a narrower electron beam energy distribution or an ion kinetic energy-dependent loss in sensitivity of the spectrometer he used \cite{dressler_energy_1985,slaughter_dissociative_2011} that we have not attempted to correct here. 

\begin{figure}
\begin{center}
\includegraphics[width=0.6\columnwidth,clip=true]{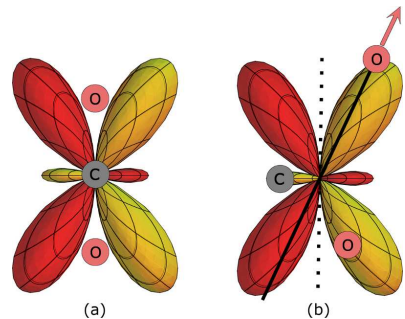}
\caption{(color online). a) One component of the doubly degenerate $^2\Pi_u$ electron attachment entrance amplitude for the 4~eV shape resonance in CO$_2$ for equilibrium geometry. b) The same entrance amplitude as Fig. \ref{EA}a, rotated with respect to the molecular axis to simulate non-axial recoil, as discussed in the text.}\label{EA}
\end{center}
\end{figure}

\begin{figure}
\begin{center}
\includegraphics[width=1.0\columnwidth,clip=true]{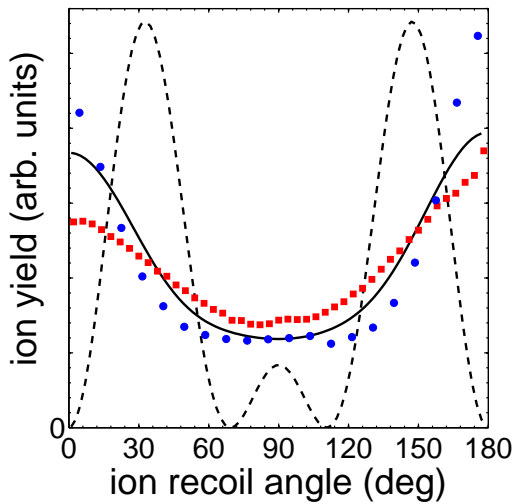}
\caption{(color online). Measured O$^-$ angular distributions from the AU (circles) and LBNL (squares) experiments, compared against the axial recoil theoretical prediction (dashed curve) and the recoil-averaged theoretical prediction (solid curve), as discussed in the text. All of the data are normalized to have the same total ion yield.}\label{AD}
\end{center}
\end{figure}

Our experimental data for the single anionic fragment is measured in a laboratory frame of reference determined by the electron beam direction. Therefore, without making any assumptions about the electron attachment dynamics, we cannot link the measurements to the molecular frame using the experimental data alone. We draw upon the quantum mechanical entrance amplitude~\cite{hax2006}, calculated from electron scattering theory, to determine the electron attachment probability as a function of the molecular orientation to make the connection. Figure \ref{EA}a displays the entrance amplitude for electron attachment for the 4~eV shape resonance at the equilibrium geometry, with C-O bond lengths of 2.2 a$_0$.  
The key features are large peaks, each about 33$^\circ$ from the O-C-O axis, small peaks orthogonal to the axis and a zero-probability for electron attachment along the axis of the molecule.  Under the axial recoil approximation (Fig. \ref{EA}a) we predict an angular distribution with respect to the incident electron directly from the squared modulus of the calculated entrance amplitude, in order to compare with the experiment. Since the experiment detects only the momentum of the atomic negative ion relative to the direction of the incident electron beam, the attachment probability is averaged over the coordinate $\phi$ azimuthal to the recoil axis to produce a laboratory frame fragment distribution as a function of $\theta$, the scattering angle of the recoil vector relative to the incident electron~\cite{hax2006}. As displayed in Fig.~\ref{AD}, there is no similarity whatsoever between the theory (dashed line) and the experiments (data points); in fact the measured angular distributions have maxima in the forward and backward directions with respect to the electron beam direction and a broad minimum in the orthogonal directions.

\section{Axial recoil breakdown}
Figure \ref{AD} illustrates the remarkable difference between the axial recoil prediction and both sets of experimental data, which is clear evidence of a breakdown of the axial recoil approximation. This is in stark contrast to our recent analysis~\cite{slaughter_dissociative_2011} of DEA through the 8.2~eV Feshbach resonance in CO$_2$, where the dissociation dynamics were successfully modeled by a simple isotropic broadening of the axial recoil approximation. In the present case it is clear that the TNI departs a from  linear geometry before the dissociation. However, an in-depth computational study of the O$^-$ and CO trajectories requires detailed knowledge of the topologies of the three relevant negative ion potential energy surfaces - having $^2\Pi_u$, $^2\Pi_g$ and $^2\Sigma_g$ symmetry at the initial geometry of neutral CO$_2$ - which is beyond the scope of the present work. Nevertheless, we can qualitatively account for the closing O-C-O bond angle by rotating the anionic fragment recoil axis, while holding the entrance amplitude fixed (see Fig. \ref{EA}b). This enables us to predict the fragment angular distribution for an O$^-$ dissociation angle fixed relative to the initial orientation of the molecular axis at the time of attachment with respect to the CO molecular axis in the molecular frame. We note that the correspondence between the molecular frame recoil angle and the bending angle is only approximate because it makes assumptions about the dissociation dynamics. By integrating over some range of deflection angles, we approximate the physical situation in which there is no single recoil axis and the fragmenting wave packet is sprayed over a range of recoil angles. In the extreme non-axial recoil case, we have averaged over all non-zero recoil angles, allowing dissociation to occur with equal likelihood at any non-linear bond angle. 

It is important to note that at the initial linear geometry, the doubly degenerate $^2\Pi_u$ resonance state gives rise to a cylindrically symmetric entrance amplitude. However, upon bending, the degeneracy is lifted and the resonance splits into non-degenerate A$'$ and A$''$ components. To produce the results for non-axial recoil, we have to assume that the DEA dynamics proceeds via the A$'$ state, as depicted in Fig.~\ref{EA}b. Bending along the perpendicular (A$''$) surface produces a qualitatively different result, with zeroes in the forward and backward direction, in poor agreement with experiment. Therefore DEA via the A$''$ state would produce zero cross section in the forward and backward directions; initial CO$_2$ orientations in which the plane of the molecule includes the momentum vector of the electron are those for which bending dynamics can fill in the zero, and the entrance amplitude to the A$''$ state is zero for these orientations. 

The experiments indicate a small but definite forward-backward asymmetry in the measured angular distributions. By calculating the entrance amplitude for slight variations in the initial CO$_2$ geometry, we found subtle but significant enhancement to the electron attachment probability for angles near the the stretched C-O bond when the opposite C-O bond is held fixed. The result is plotted in Fig.~\ref{AD} for the extreme non-axial recoil treatment.  In contrast to the axial recoil prediction, there are no narrow peaks at 33$^\circ$ and 90$^\circ$ recoil angles, but a broad minimum at 90$^\circ$ with smoothly increasing ion yields towards 0$^\circ$ and 180$^\circ$.  When one C-O bond is stretched, an increase of about 10\% is found in the calculated backward ion yield, compared to the yield around 0$^\circ$, which appears to be an underestimation when compared with the experimental results. This forward-backward asymmetry in the ion yield suggests that DEA is favored by attachment at asymmetric geometries, via the variation of the entrance amplitude or survival probability with respect to the geometry at attachment. The good agreement between the extreme non-axial recoil prediction with the experimental data indicates that the final O-C-O bond angle of the dissociating TNI falls within a broad range and is likely to be dependent upon the exact point of origin of the wavepacket on the $^2\Pi_u$ PES. 

\section{Dissociation mechanisms}

To gain further insight into the nature of the post-attachment dissociation dynamics, we carried out a series of electronic structure calculations using the Columbus program~\cite{COLUMBUS}, more extensive than those undertaken in our previous study~\cite{slaughter_dissociative_2011} of DEA through the 8.2 eV resonance. The present calculations employed Dunning's aug-cc-pvtz basis~\cite{dunning1989, kendall1992}, augmented with two additional diffuse s- and two diffuse p-functions on the carbon and oxygens. We carried out multi-reference configuration-interaction (CAS plus singles and doubles CI) calculations using natural orbitals derived from multi-configuration self-consistent field calculations on the relevant anion states.  The first five molecular orbitals (MO's)were constrained to be doubly occupied.

In addition to the previously mentioned conical intersection between the $^2\Pi$ shape- and Feshbach resonances near asymmetric linear geometry, we found two seams of symmetry conical intersections, at bent C$_{\rm{2v}}$ geometries where the CO bond distances are equal, between the A$'$ and A$''$ components of the shape- and Feshbach anion states. (Recall that upon bending, the doubly degenerate $\Pi$ states split into non-degenerate (A$'$ and A$''$) components.) The energy minima along both the A$'$ and A$''$ conical intersection seams occur at  similar geometries where the OCO bond angle is approximately 130 degrees and the CO bond distances are near 2.4 bohr.

A third conical intersection occurs near linear symmetrically stretched geometry between the $^2$A$'$ component of the shape resonance and the $^2$A$'$ ( $^2\Sigma^+$) anion state, which becomes a virtual state at CO distances near the equilibrium geometry of neutral CO$_2$~\cite{somm2004}. To determine which of these conical intersections might play a role in the dissociation dynamics, we must also appeal to earlier theoretical and experimental work on electron-CO$_2$ scattering in addition to the presently observed angular distributions.
\subsection{Virtual-shape conical intersection}
\begin{figure}
\begin{center}
\includegraphics[width=0.95\columnwidth]{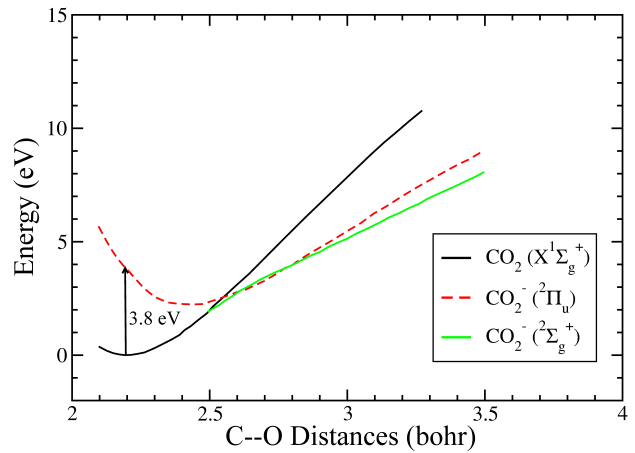}
\caption{(color online). Potential curves of neutral CO$_2$ and the $^2\Pi_u$ and $^2\Sigma_{u}^+$ anion states from CI singles and doubles calculations in linear geometry, with the two CO distances held equal. The neutral curve was shifted by $\sim$1.7 eV relative to the anion curves to place the shape resonance at 3.8 eV at the equilibrium geometry of CO$_2$. The anion curves cross near 2.6 and 2.8 bohr, indicating a seam of conical intersection close to linear geometry.}\label{shape_virt}
\end{center}
\end{figure}

DEA near 4 eV is initiated by electron capture by neutral CO$_2$ into the $^2\Pi_u$ shape resonance near linear equilibrium geometry (R$_{\rm{CO}}$=2.1944 bohr). Consider first dissociation via the virtual state/shape resonance conical intersection. Our calculations, which are displayed in Fig.~\ref{shape_virt},  place this near symmetric linear geometry with R$_{\rm{CO}}\approx$ 2.6 bohr. The path from neutral equilibrium geometry to the stretched linear geometry is slightly exothermic. A wavepacket created on the $^2\Pi_u$ surface could reach the conical intersection with the $^2\Sigma_{u}^+$ state on a linear path with no barriers and then dissociate to ground state CO + O$^-$ on the lower $^2$A$'$ surface. Since the lower  $^2$A$'$ surface near the conical intersection is relatively flat with respect to bending, the wavepacket would spread rapidly on the lower surface and the dissociation with respect to bending would be statistical, producing an angular distribution consistent with the extreme non-axial recoil case discussed in the previous section. The fact that boomerang structure is observed in vibrational excitation indicates that symmetric stretching motion during the lifetime of the resonance reaches the outer turning point~\cite{mccurdy_resonant_2003} near the conical intersection with the virtual state in Fig.~\ref{shape_virt}.  However, a mechanism for DEA through this resonance must also be consistent with the observed vibrational distribution of the products.   The 4 eV resonance produces CO mostly in $\nu = 0$, 1, and 2 with up to $\nu = 4$ detectable~\cite{dressler_energy_1985,stamatovic_vibrational_1973}.  Although the conical intersection occurs with both CO bonds stretched to R $>$ 2.5 bohr, the gradient to leave the intersection (the g vector in the standard analysis) involves asymmetric stretch.  It is therefore possible that asymmetric stretching motion could end up depositing much of the vibrational energy into dissociation.  We cannot eliminate the possibility that this mechanism would produce the observed vibrational excitation without calculating the full dissociation dynamics on the two coupled potential surfaces.  So this intersection may be involved in DEA via the 4 eV resonance. 

\subsection{Symmetry allowed conical intersections}
We turn next to the symmetry conical intersections between the pairs of A$'$ and A$''$ components of the shape and Feshbach resonances, which occur at stretched and bent geometries where they are electronically bound. To understand the dynamics, we need to know something about the resonance width (inverse lifetimes) of the shape resonance and its dependence on geometry. From our previous studies of resonant vibrational excitation~\cite{resc2002,mccurdy_resonant_2003}, we know that while the energy of the A$'$ component of the shape resonance decreases upon bending, its width (at CO distances where it is electronically unbound) sharply increases upon bending owing to the addition of an s-wave component into the wave function. This leads to rapid autodetachment. By contrast, the width of the A$''$ component is relatively insensitive to geometry. But we know from the results of the previous section that the observed angular distributions are completely inconsistent with attachment on an A$''$ surface, which is expected to produce minima in the forward and backward directions. The reason that dissociation on the A$''$ surface does not occur, as we learned from our structure calculations, is that there is a barrier to dissociation on the A$''$ surface, as its energy increases in C$_{\rm{2v}}$ symmetry upon bending. This observation points to a dissociation path around the conical intersection on the lower cone of the A$'$ resonance surface. Because of the rapid increase of the resonance width upon bending near the initial geometry, the dissociation path would require an initial symmetric stretch motion in linear geometry, followed by bending motion around the conical intersection and then asymmetric dissociation to products.While such a mechanism cannot be ruled out on energetic grounds, it can be expected to be accompanied by significant production of vibrationally excited CO products, which as we have noted above is not observed.

\begin{figure}
\begin{center}
\includegraphics[width=0.8\columnwidth]{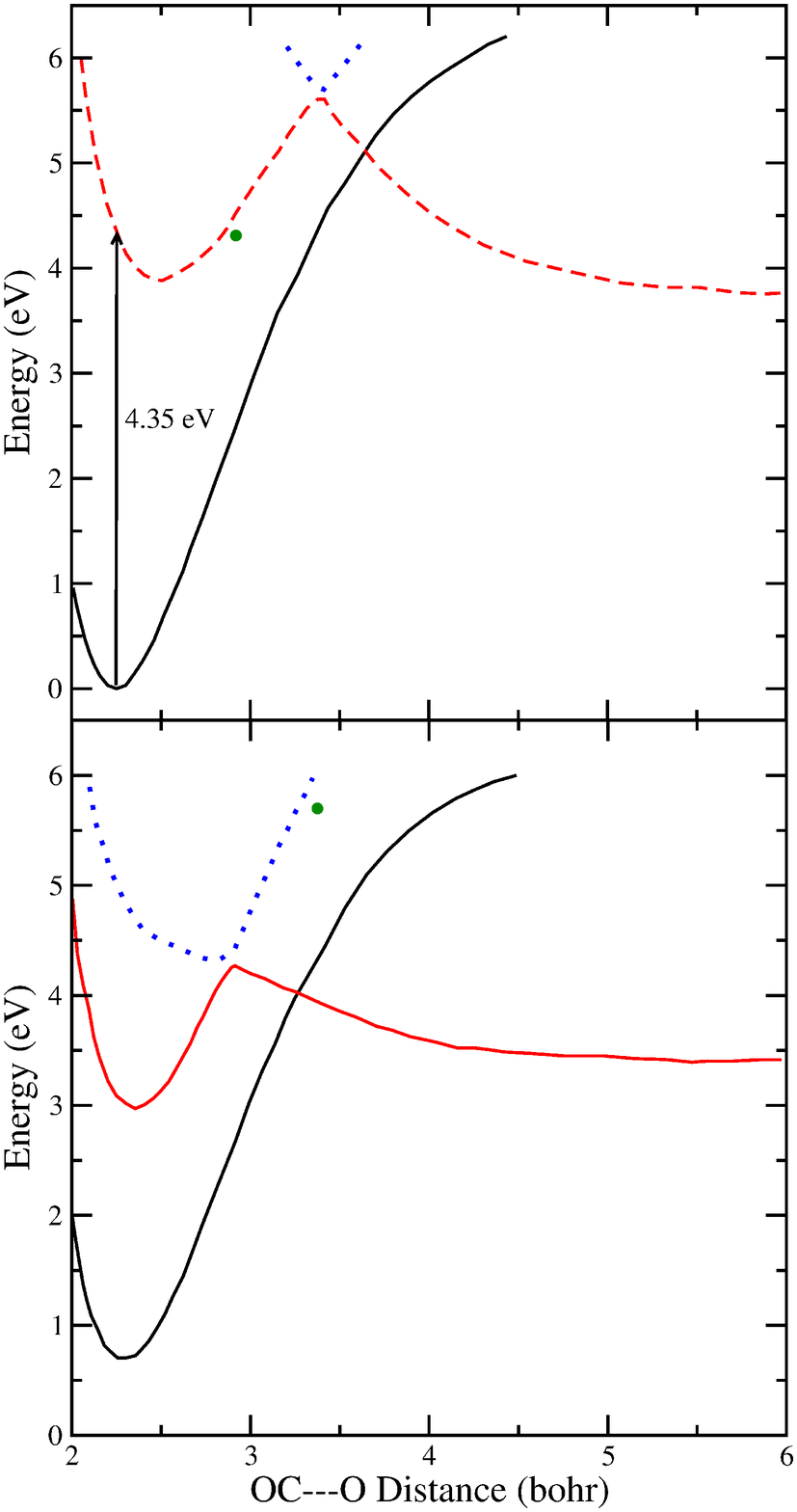}
\caption{(color online). Potential curves of neutral CO$_2$ and the shape- and Feshbach anion states, showing the avoided crossing between the two resonance states. No adjustment of the neutral and anion curves was made. Upper: Linear geometry; the point near 3 bohr shows the position of the crossing when the OCO angle is 140 degrees. Lower: O-C-O angle is 140 degrees; the point near 3.4 bohr shows the position of the crossing when the OCO angle is 180 degrees.}\label{curves}
\end{center}
\end{figure}
\subsection{Accidental conical intersection}
These observations led us to reconsider the conical intersection between shape- and Feshbach resonances in near-linear, asymmetrically stretched geometry which we had previously studied in connection with DEA at 8.2 eV. This is a conical intersection not determined through symmetry, i.e., an accidental conical intersection. Those earlier calculations placed the crossing of the shape and Feshbach anion states at a geometry with the stretched CO distance near 3.8 bohr, when the other CO distance was fixed at 2.1944 bohr. The present calculations were done with a larger basis (Rydberg augmented triple-zeta vs. our earlier double-zeta basis with a fewer Rydberg functions) and a larger active space ( two $\sigma^*$ vs one $\sigma^*$ orbitals). More importantly, the molecular orbitals used in the CAS-CI were obtained from MCSCF calculations on the $^2\Pi$ anion states, rather than our earlier state-averaged MCSCF calculations on the neutral target. The present calculations place the linear crossing near 3.2 (vs our earlier 3.8) bohr, at an energy $\sim$1.3 eV above the initial CO$_2$ neutral geometry. Bending with respect to this linear geometry beyond O-C-O angles of $\sim$150 degrees removes the barrier to dissociation, and can consequently result in cold CO + O$^-$ products.  The combination of initial asymmetric stretch motion in linear geometry, which minimizes the probability of autodetachment, followed by bending around the conical intersection on the lower A$'$ surface, can thus explain the measured angular distributions and the experimental observation of a relatively cold CO product. The barrier to dissociation in linear geometry may also explain the observation that the onset of DEA  via the low energy shape resonance is not truly vertical~\cite{stamatovic1973}. 

\subsection{Confirmation of accidental conical intersection mechanism}
To confirm the proposed mechanism, we undertook a final set of electronic structure calculations, using an entirely different prescription for choosing the molecular orbital basis. This prescription was designed for balance of the N and N+1 electron systems and an accurate description of many asymptotic states of CO + O, neutral and anion. We began by carrying out  state-averaged MCSCF calculations on the fragment CO and O  species, averaging over both neutral and anion states, to produce a basis of nineteen MO's. The oxygen states included in the average were O $^1$S, $^1$D, $^3$P, $^5$S and O$^-$ $^2$P, while the diatomic state average included CO $^1\Sigma^+$, $^3\Sigma^+$, $^3\Sigma^-$, $^3\Pi$, $^3\Delta$ and CO$^-$ $^2\Pi$, $^4\Pi$, $^4\Sigma^-$ and $^2\Sigma^+$. We then performed  MCSCF calculations on neutral CO$_2$ in this restricted space of nineteen orbitals to obtain twelve active orbitals as linear combinations of the CO and O MCSCF orbital bases. The final CI calculations had an active space of nineteen orbitals, the first three of which were restricted to be doubly occupied in all configurations. From the remaining active orbitals, a large number of reference configurations was generated, followed by all single excitations into the remaining space of virtual orbitals. This prescription resulted in $\sim$24 million configurations for the anion states and $\sim$7 million configurations for the neutral states.  

The results are graphically displayed in Fig.~\ref{curves}.  These calculations also show that the barrier to dissociation in linear geometry of $\sim$ 1.3 eV  is removed when the O-C-O angle is bent from linear geometry to $\sim140$ degrees.

\section{Conclusions}
The present study combines experimental data along with theoretical analysis of dissociative electron attachment to carbon dioxide via the 4 eV $^2\Pi_u$ shape resonance. We have demonstrated that an understanding of anion dissociation dynamics beyond simple one-dimensional models is crucial in interpreting the measured angular distributions. Although several possible dissociation mechanisms involving conical intersections have been identified and discussed, the most likely scenario points to an initial linear asymmetric stretch motion  to geometries where the autodetachment probability is small, followed by bending motion around a conical intersection and dissociation to produce ground-state CO + O$^-$. The proposed mechanism is consistent with the observed non-axial recoil angular distributions and the observation of relatively cold CO product. It is also possible that the conical intersection in symmetric stretch between the shape resonance and virtual states could contribute to dissociative attachment through the shape resonance, since earlier calculations indicate that the dynamics on the negative ion potential surface visits this region of configuration space, but without a full treatment of dynamics of passage through the conical intersection we cannot verify that this mechanism is consistent with the observed angular and vibrational distributions of products. This work, when combined with our recent analyses of the 8~eV Feshbach resonance~\cite{slaughter_dissociative_2011, MoradmandCO28eV}, provides the most comprehensive picture of the metastable anions of CO$_2$ to date.
\begin{acknowledgments}
Work at University of California Lawrence Berkeley National Laboratory performed under the auspices of the US Department of Energy DOE) under Contract DE-AC02-05CH11231 and was supported by the U.S. DOE Office of Basic Energy Sciences, Division of Chemical Sciences.  Work at Auburn University performed under DOE contract DE-FG02-10ER16146. Work at Temple University performed under DOE contract DE-FG02-08ER15983. 
\end{acknowledgments}

\bibliography{DEA-CO2-shape}

\begin{thebibliography}{26}
\expandafter\ifx\csname natexlab\endcsname\relax\def\natexlab#1{#1}\fi
\expandafter\ifx\csname bibnamefont\endcsname\relax
  \def\bibnamefont#1{#1}\fi
\expandafter\ifx\csname bibfnamefont\endcsname\relax
  \def\bibfnamefont#1{#1}\fi
\expandafter\ifx\csname citenamefont\endcsname\relax
  \def\citenamefont#1{#1}\fi
\expandafter\ifx\csname url\endcsname\relax
  \def\url#1{\texttt{#1}}\fi
\expandafter\ifx\csname urlprefix\endcsname\relax\def\urlprefix{URL }\fi
\providecommand{\bibinfo}[2]{#2}
\providecommand{\eprint}[2][]{\url{#2}}

\bibitem[{\citenamefont{Boness and Schulz}(1974)}]{Bone1969}
\bibinfo{author}{\bibfnamefont{M.~J.~W.} \bibnamefont{Boness}}
  \bibnamefont{and} \bibinfo{author}{\bibfnamefont{G.~J.}
  \bibnamefont{Schulz}}, \bibinfo{journal}{Physical Review A}
  \textbf{\bibinfo{volume}{9}}, \bibinfo{pages}{1969} (\bibinfo{year}{1974}).

\bibitem[{\citenamefont{Morrison et~al.}(1977)\citenamefont{Morrison, Lane, and
  Collins}}]{Morrison1977}
\bibinfo{author}{\bibfnamefont{M.~A.} \bibnamefont{Morrison}},
  \bibinfo{author}{\bibfnamefont{N.~F.} \bibnamefont{Lane}}, \bibnamefont{and}
  \bibinfo{author}{\bibfnamefont{L.~A.} \bibnamefont{Collins}},
  \bibinfo{journal}{Phys. Rev. A} \textbf{\bibinfo{volume}{15}},
  \bibinfo{pages}{2186} (\bibinfo{year}{1977}).

\bibitem[{\citenamefont{Allan}(2001)}]{allan_selectivity_2001}
\bibinfo{author}{\bibfnamefont{M.}~\bibnamefont{Allan}},
  \bibinfo{journal}{Physical Review Letters} \textbf{\bibinfo{volume}{87}},
  \bibinfo{pages}{033201} (\bibinfo{year}{2001}).

\bibitem[{\citenamefont{{T. N. Rescigno, W. A. Isaacs, A. E. Orel, H. -D. Meyer
  and C. W. McCurdy}}(2002)}]{resc2002}
\bibinfo{author}{\bibnamefont{{T. N. Rescigno, W. A. Isaacs, A. E. Orel, H. -D.
  Meyer and C. W. McCurdy}}}, \bibinfo{journal}{Phys. Rev. A}
  \textbf{\bibinfo{volume}{65}}, \bibinfo{pages}{032716}
  (\bibinfo{year}{2002}).

\bibitem[{\citenamefont{{McCurdy} et~al.}(2003)\citenamefont{{McCurdy}, Isaacs,
  Meyer, and Rescigno}}]{mccurdy_resonant_2003}
\bibinfo{author}{\bibfnamefont{C.}~\bibnamefont{{McCurdy}}},
  \bibinfo{author}{\bibfnamefont{W.}~\bibnamefont{Isaacs}},
  \bibinfo{author}{\bibfnamefont{H.-D.} \bibnamefont{Meyer}}, \bibnamefont{and}
  \bibinfo{author}{\bibfnamefont{T.}~\bibnamefont{Rescigno}},
  \bibinfo{journal}{Physical Review A} \textbf{\bibinfo{volume}{67}},
  \bibinfo{pages}{042708} (\bibinfo{year}{2003}).

\bibitem[{\citenamefont{Vanroose et~al.}(2004)\citenamefont{Vanroose, Zhang,
  McCurdy, and Rescigno}}]{vanr2004}
\bibinfo{author}{\bibfnamefont{W.}~\bibnamefont{Vanroose}},
  \bibinfo{author}{\bibfnamefont{Z.}~\bibnamefont{Zhang}},
  \bibinfo{author}{\bibfnamefont{C.~W.} \bibnamefont{McCurdy}},
  \bibnamefont{and} \bibinfo{author}{\bibfnamefont{T.~N.}
  \bibnamefont{Rescigno}}, \bibinfo{journal}{Phys. Rev. Lett.}
  \textbf{\bibinfo{volume}{92}}, \bibinfo{pages}{053201}
  (\bibinfo{year}{2004}).

\bibitem[{\citenamefont{Allan}(2002)}]{allan_vibrational_2002}
\bibinfo{author}{\bibfnamefont{M.}~\bibnamefont{Allan}}, \bibinfo{journal}{J.
  Phys. B} \textbf{\bibinfo{volume}{35}}, \bibinfo{pages}{L387}
  (\bibinfo{year}{2002}).

\bibitem[{\citenamefont{{T. Sommerfeld, H. -D. Meyer and L. S.
  Cederbaum}}(2004)}]{somm2004}
\bibinfo{author}{\bibnamefont{{T. Sommerfeld, H. -D. Meyer and L. S.
  Cederbaum}}}, \bibinfo{journal}{Phys. Chem. Chem. Phys.}
  \textbf{\bibinfo{volume}{6}}, \bibinfo{pages}{42} (\bibinfo{year}{2004}).

\bibitem[{\citenamefont{Chantry}(1972)}]{chant1972}
\bibinfo{author}{\bibfnamefont{P.~J.} \bibnamefont{Chantry}},
  \bibinfo{journal}{J. Chem. Phys.} \textbf{\bibinfo{volume}{57}},
  \bibinfo{pages}{3180} (\bibinfo{year}{1972}).

\bibitem[{\citenamefont{Sizun and Goursaud}(1979)}]{sizun1979}
\bibinfo{author}{\bibfnamefont{M.}~\bibnamefont{Sizun}} \bibnamefont{and}
  \bibinfo{author}{\bibfnamefont{S.}~\bibnamefont{Goursaud}},
  \bibinfo{journal}{J. Chem. Phys.} \textbf{\bibinfo{volume}{71}},
  \bibinfo{pages}{4042} (\bibinfo{year}{1979}).

\bibitem[{\citenamefont{Dressler and Allan}(1985)}]{dressler_energy_1985}
\bibinfo{author}{\bibfnamefont{R.}~\bibnamefont{Dressler}} \bibnamefont{and}
  \bibinfo{author}{\bibfnamefont{M.}~\bibnamefont{Allan}},
  \bibinfo{journal}{Chemical Physics} \textbf{\bibinfo{volume}{92}},
  \bibinfo{pages}{449} (\bibinfo{year}{1985}).

\bibitem[{\citenamefont{Slaughter et~al.}(2011)\citenamefont{Slaughter,
  Adaniya, Rescigno, Haxton, Orel, {McCurdy}, and
  Belkacem}}]{slaughter_dissociative_2011}
\bibinfo{author}{\bibfnamefont{D.~S.} \bibnamefont{Slaughter}},
  \bibinfo{author}{\bibfnamefont{H.}~\bibnamefont{Adaniya}},
  \bibinfo{author}{\bibfnamefont{T.~N.} \bibnamefont{Rescigno}},
  \bibinfo{author}{\bibfnamefont{D.~J.} \bibnamefont{Haxton}},
  \bibinfo{author}{\bibfnamefont{A.~E.} \bibnamefont{Orel}},
  \bibinfo{author}{\bibfnamefont{C.~W.} \bibnamefont{{McCurdy}}},
  \bibnamefont{and} \bibinfo{author}{\bibfnamefont{A.}~\bibnamefont{Belkacem}},
  \bibinfo{journal}{J. Phys. B} \textbf{\bibinfo{volume}{44}},
  \bibinfo{pages}{205203} (\bibinfo{year}{2011}).

\bibitem[{\citenamefont{Wu et~al.}(2012)\citenamefont{Wu, Xia, Wang, Li, Zeng,
  and Tian}}]{wu_renner-teller_2012}
\bibinfo{author}{\bibfnamefont{B.}~\bibnamefont{Wu}},
  \bibinfo{author}{\bibfnamefont{L.}~\bibnamefont{Xia}},
  \bibinfo{author}{\bibfnamefont{Y.-F.} \bibnamefont{Wang}},
  \bibinfo{author}{\bibfnamefont{H.-K.} \bibnamefont{Li}},
  \bibinfo{author}{\bibfnamefont{X.-J.} \bibnamefont{Zeng}}, \bibnamefont{and}
  \bibinfo{author}{\bibfnamefont{S.~X.} \bibnamefont{Tian}},
  \bibinfo{journal}{Phys. Rev. A} \textbf{\bibinfo{volume}{85}},
  \bibinfo{pages}{052709} (\bibinfo{year}{2012}).

\bibitem[{\citenamefont{Stamatovic and Schultz}(1973)}]{stamatovic1973}
\bibinfo{author}{\bibfnamefont{A.}~\bibnamefont{Stamatovic}} \bibnamefont{and}
  \bibinfo{author}{\bibfnamefont{G.~J.} \bibnamefont{Schultz}},
  \bibinfo{journal}{Phys. Rev. A.} \textbf{\bibinfo{volume}{7}},
  \bibinfo{pages}{589} (\bibinfo{year}{1973}).

\bibitem[{\citenamefont{Orient and Srivastava}(1983)}]{Orie1983}
\bibinfo{author}{\bibfnamefont{O.~J.} \bibnamefont{Orient}} \bibnamefont{and}
  \bibinfo{author}{\bibfnamefont{S.~K.} \bibnamefont{Srivastava}},
  \bibinfo{journal}{Chemical Physics Letters} \textbf{\bibinfo{volume}{96}},
  \bibinfo{pages}{681 } (\bibinfo{year}{1983}).

\bibitem[{\citenamefont{Haxton et~al.}(2011)\citenamefont{Haxton, Adaniya,
  Slaughter, Rudek, Osipov, Weber, Rescigno, {McCurdy}, and
  Belkacem}}]{haxton_observation_2011}
\bibinfo{author}{\bibfnamefont{D.~J.} \bibnamefont{Haxton}},
  \bibinfo{author}{\bibfnamefont{H.}~\bibnamefont{Adaniya}},
  \bibinfo{author}{\bibfnamefont{D.~S.} \bibnamefont{Slaughter}},
  \bibinfo{author}{\bibfnamefont{B.}~\bibnamefont{Rudek}},
  \bibinfo{author}{\bibfnamefont{T.}~\bibnamefont{Osipov}},
  \bibinfo{author}{\bibfnamefont{T.}~\bibnamefont{Weber}},
  \bibinfo{author}{\bibfnamefont{T.~N.} \bibnamefont{Rescigno}},
  \bibinfo{author}{\bibfnamefont{C.~W.} \bibnamefont{{McCurdy}}},
  \bibnamefont{and} \bibinfo{author}{\bibfnamefont{A.}~\bibnamefont{Belkacem}},
  \bibinfo{journal}{Phys. Rev. A} \textbf{\bibinfo{volume}{84}},
  \bibinfo{pages}{030701R} (\bibinfo{year}{2011}).

\bibitem[{\citenamefont{Haxton et~al.}(2006)\citenamefont{Haxton, McCurdy, and
  Rescigno}}]{hax2006}
\bibinfo{author}{\bibfnamefont{D.~J.} \bibnamefont{Haxton}},
  \bibinfo{author}{\bibfnamefont{C.~W.} \bibnamefont{McCurdy}},
  \bibnamefont{and} \bibinfo{author}{\bibfnamefont{T.~N.}
  \bibnamefont{Rescigno}}, \bibinfo{journal}{Phys. Rev. A}
  \textbf{\bibinfo{volume}{73}}, \bibinfo{pages}{062724}
  (\bibinfo{year}{2006}).

\bibitem[{\citenamefont{Moradmand
  et~al.}(2013{\natexlab{a}})\citenamefont{Moradmand, Williams, Landers, and
  Fogle}}]{moradmand_momentum-imaging_2013}
\bibinfo{author}{\bibfnamefont{A.}~\bibnamefont{Moradmand}},
  \bibinfo{author}{\bibfnamefont{J.~B.} \bibnamefont{Williams}},
  \bibinfo{author}{\bibfnamefont{A.~L.} \bibnamefont{Landers}},
  \bibnamefont{and} \bibinfo{author}{\bibfnamefont{M.}~\bibnamefont{Fogle}},
  \bibinfo{journal}{Review of Scientific Instruments}
  \textbf{\bibinfo{volume}{84}}, \bibinfo{pages}{033104}
  (\bibinfo{year}{2013}{\natexlab{a}}).

\bibitem[{\citenamefont{Adaniya et~al.}(2012)\citenamefont{Adaniya, Slaughter,
  Osipov, Weber, and Belkacem}}]{adaniya_momentum_2012}
\bibinfo{author}{\bibfnamefont{H.}~\bibnamefont{Adaniya}},
  \bibinfo{author}{\bibfnamefont{D.~S.} \bibnamefont{Slaughter}},
  \bibinfo{author}{\bibfnamefont{T.}~\bibnamefont{Osipov}},
  \bibinfo{author}{\bibfnamefont{T.}~\bibnamefont{Weber}}, \bibnamefont{and}
  \bibinfo{author}{\bibfnamefont{A.}~\bibnamefont{Belkacem}},
  \bibinfo{journal}{Rev. Sci. Instr.} \textbf{\bibinfo{volume}{83}},
  \bibinfo{pages}{023106} (\bibinfo{year}{2012}).

\bibitem[{\citenamefont{Ramsey}(1985)}]{ramsey_molecular_1985}
\bibinfo{author}{\bibfnamefont{N.}~\bibnamefont{Ramsey}},
  \emph{\bibinfo{title}{Molecular Beams}} (\bibinfo{publisher}{Clarendon Press
  ; Oxford University Press}, \bibinfo{address}{Oxford; New York},
  \bibinfo{year}{1985}), ISBN \bibinfo{isbn}{0198520212 9780198520214}.

\bibitem[{\citenamefont{Chantry and Schulz}(1967)}]{chan1967}
\bibinfo{author}{\bibfnamefont{P.~J.} \bibnamefont{Chantry}} \bibnamefont{and}
  \bibinfo{author}{\bibfnamefont{G.~J.} \bibnamefont{Schulz}},
  \bibinfo{journal}{Phys. Rev.} \textbf{\bibinfo{volume}{156}},
  \bibinfo{pages}{134} (\bibinfo{year}{1967}).

\bibitem[{\citenamefont{Lischka et~al.}(2012)\citenamefont{Lischka, Shepard,
  Shavitt, Pitzer, Dallos, M\"uller, Szalay, Brown, Ahlrichs, B\"ohm
  et~al.}}]{COLUMBUS}
\bibinfo{author}{\bibfnamefont{H.}~\bibnamefont{Lischka}},
  \bibinfo{author}{\bibfnamefont{R.}~\bibnamefont{Shepard}},
  \bibinfo{author}{\bibfnamefont{I.}~\bibnamefont{Shavitt}},
  \bibinfo{author}{\bibfnamefont{R.~M.} \bibnamefont{Pitzer}},
  \bibinfo{author}{\bibfnamefont{M.}~\bibnamefont{Dallos}},
  \bibinfo{author}{\bibfnamefont{T.}~\bibnamefont{M\"uller}},
  \bibinfo{author}{\bibfnamefont{P.~G.} \bibnamefont{Szalay}},
  \bibinfo{author}{\bibfnamefont{F.~B.} \bibnamefont{Brown}},
  \bibinfo{author}{\bibfnamefont{R.}~\bibnamefont{Ahlrichs}},
  \bibinfo{author}{\bibfnamefont{H.~J.} \bibnamefont{B\"ohm}},
  \bibnamefont{et~al.}, \emph{\bibinfo{title}{Columbus, an ab initio electronic
  structure program, release 7.0}} (\bibinfo{year}{2012}).

\bibitem[{\citenamefont{Dunning}(1981)}]{dunning1989}
\bibinfo{author}{\bibfnamefont{T.~H.} \bibnamefont{Dunning}},
  \bibinfo{journal}{J. Chem. Phys.} \textbf{\bibinfo{volume}{90}},
  \bibinfo{pages}{1007} (\bibinfo{year}{1981}).

\bibitem[{\citenamefont{Kendall et~al.}(1992)\citenamefont{Kendall, Dunning,
  and Harrison}}]{kendall1992}
\bibinfo{author}{\bibfnamefont{R.~A.} \bibnamefont{Kendall}},
  \bibinfo{author}{\bibfnamefont{T.~H.} \bibnamefont{Dunning}},
  \bibnamefont{and} \bibinfo{author}{\bibfnamefont{R.~J.}
  \bibnamefont{Harrison}}, \bibinfo{journal}{J. Chem. Phys.}
  \textbf{\bibinfo{volume}{96}}, \bibinfo{pages}{6796} (\bibinfo{year}{1992}).

\bibitem[{\citenamefont{Stamatovic and
  Schulz}(1973)}]{stamatovic_vibrational_1973}
\bibinfo{author}{\bibfnamefont{A.}~\bibnamefont{Stamatovic}} \bibnamefont{and}
  \bibinfo{author}{\bibfnamefont{G.}~\bibnamefont{Schulz}},
  \bibinfo{journal}{Physical Review A} \textbf{\bibinfo{volume}{7}},
  \bibinfo{pages}{589} (\bibinfo{year}{1973}).

\bibitem[{\citenamefont{Moradmand
  et~al.}(2013{\natexlab{b}})\citenamefont{Moradmand, Slaughter, Landers, and
  Fogle}}]{MoradmandCO28eV}
\bibinfo{author}{\bibfnamefont{A.}~\bibnamefont{Moradmand}},
  \bibinfo{author}{\bibfnamefont{D.~S.} \bibnamefont{Slaughter}},
  \bibinfo{author}{\bibfnamefont{A.~L.} \bibnamefont{Landers}},
  \bibnamefont{and} \bibinfo{author}{\bibfnamefont{M.}~\bibnamefont{Fogle}},
  \bibinfo{journal}{submitted}  (\bibinfo{year}{2013}{\natexlab{b}}).

\end{thebibliography}

\end{document}